# A Macroscopic Theory of Saturated Ferromagnetic Conductors

Jiashi Yang (jyang1@unl.edu)
Department of Mechanical and Materials Engineering
University of Nebraska-Lincoln, Lincoln, NE 68588-0526, USA

**Abstract** - A phenomenological theory of rigid and saturated ferromagnetic conductors is constructed from a four-continuum model consisting of a rigid lattice continuum, a bound charge continuum for polarization, a circulating current continuum for magnetization, and a free charge continuum for electrical conduction. The basic laws of physics are applied to the four continua. Thermal couplings and the related dissipative effects are also included. The theory includes the Landau–Lifshitz–Gilbert equation as one of a system of simultaneous equations.

**Keywords**: ferromagnetic; conductor; saturation; LL equation; LLG equation

## 1. Four-Continuum Model

Ferromagnets may be insulators [1-10] or conductors [11]. The nonlinear and macroscopic theories for ferromagnets [1-10], rigid or elastic, are for insulators. In this paper we develop a macroscopic theory of rigid and saturated ferromagnets using the multi-continuum model constructed by Tiersten in [1,5,12-15]. The basic behaviors of ferromagnetic metals can be described by the four interpenetrating and interacting continua shown in Figs. 1-4. They are the lattice continuum, bound charge continuum, circulating current or spin continuum, and free charge fluid. We use the Cartesian tensor notation. The spatial coordinates are written as $x_k$ or **x**. $V$, $S$ and $C$ represent volumes, surfaces and curves fixed in space. The outward unit normal of $S$ is **n**.

The lattice continuum in Fig. 1 is rigid and stationary. Its charge density is $\mu^l(\mathbf{x})$. The lattice continuum is under the usual mechanical surface traction **t** and mechanical body force **f**. $\mu^l(\mathbf{x})$ experiences a force under the Maxwellian electric field **E**. The lattice continuum interacts with the bound charge continuum with an effective electric field $\mathbf{E}^b$. It interacts with the spin continuum with a local force $\mathbf{f}^L$ and a local couple $\mathbf{c}^L$ produced by a local magnetic induction $\mathbf{B}^L$. The lattice continuum interacts with the free charge fluid with an effective electric field $\mathbf{E}^e$.

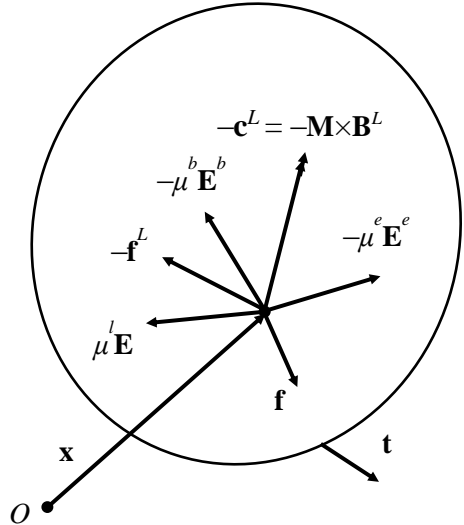

Fig. 1. Forces and couples on the lattice continuum.

The bound charge continuum in Fig. 2 is massless and carries a negative charge density $\mu^b$. $\mu^b$ experiences a force under the Maxwellian electric field **E** and magnetic induction **B**. $\mu^b$ can displace a little from the lattice continuum by an infinitesimal displacement field **η**(**x**,*t*). At the reference state **η**=0. **η** preserves the volume of the bound charge continuum, i.e.,

$$\eta_{k,k} = \frac{\partial \eta_k}{\partial x_k} = \nabla \cdot \mathbf{\eta} = 0. \tag{1.1}$$

The sum of the lattice charge density $\mu^l$ at **x** and the bound charge density $\mu^b$ at **x**+**η** is the lattice residual charge density $\mu^r$:

$$\mu^l(\mathbf{x}) + \mu^b(\mathbf{x}+\mathbf{\eta}) = \mu^r(\mathbf{x}). \tag{1.2}$$

In terms of **η**, the polarization per unit volume is defined by

$$\mathbf{P} = \mu^l(\mathbf{x})(-\mathbf{\eta}) = \mu^b(\mathbf{x}+\mathbf{\eta})\mathbf{\eta} \cong \mu^b(\mathbf{x})\mathbf{\eta}. \tag{1.3}$$

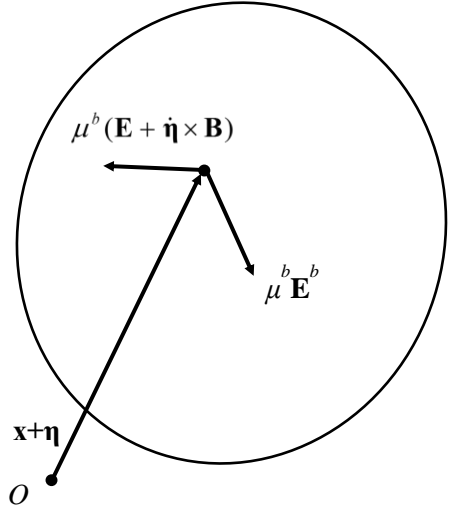

Fig. 2. Forces on the bound charge continuum.

The spin continuum in Fig. 3 is massless. It carries distributed circulating currents which produce a magnetic moment field **M**(**x**,*t*) per unit volume [16]. **M** cannot move relative to the lattice but it can rotate with respect to the lattice. It experiences a force $\mathbf{f}^M$ and a couple $\mathbf{c}^M$ under the Maxwellian magnetic induction **B** through

$$\mathbf{c}^L = \mathbf{M} \times \mathbf{B}^L, \tag{1.4}$$
$$\mathbf{c}^M = \mathbf{M} \times \mathbf{B}. \tag{1.5}$$

The spin continuum also experiences a distributed couple **M**×**F** per unit area on its boundary surface due to an effective exchange field **F**. Since **F** and $\mathbf{B}^L$ act on **M** through cross products, without loss of generality, it can be assumed that [1]

$$\mathbf{F} \cdot \mathbf{M} = 0, \tag{1.6}$$
$$\mathbf{B}^L \cdot \mathbf{M} = 0. \tag{1.7}$$

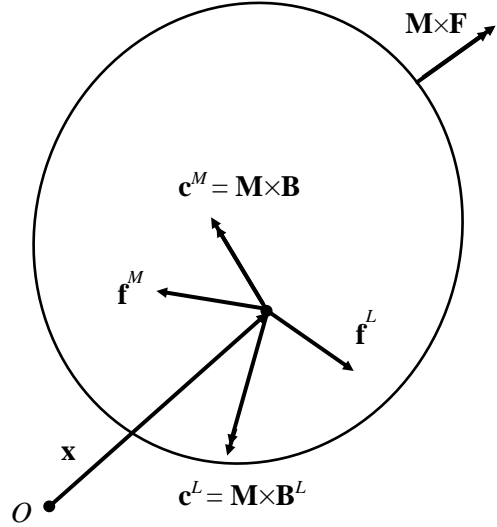

Fig. 3. Forces and couples on the spin continuum.

The free charge fluid in Fig. 4 is also massless. Its charge density is $\mu^e(\mathbf{x},t)$ and its velocity field is $\mathbf{v}^e(\mathbf{x},t)$. It experiences a force under the Maxwellian electric field $\mathbf{E}$ and magnetic induction $\mathbf{B}$. The continuity equation for the free charge is

$$\frac{\partial \mu^e}{\partial t} + (\mu^e v_k^e)_{,k} = 0. \tag{1.8}$$

We denote the total charge density and the current density by

$$\mu = \mu^r + \mu^e, \quad \mathbf{J} = \mu^e \mathbf{v}^e, \tag{1.9}$$

which satisfy

$$\frac{\partial \mu}{\partial t} + J_{k,k} = 0. \tag{1.10}$$

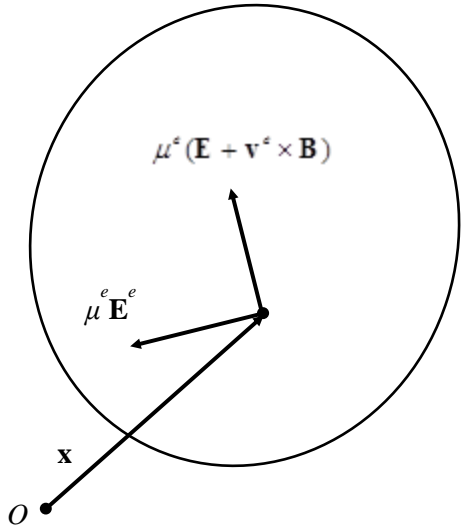

Fig. 4. Forces on the free charge fluid.

Gauss' law for the electric field can be written as
$$\oint_S \varepsilon_0 \mathbf{n} \cdot \mathbf{E} dS = \int_V [\mu^l(\mathbf{x}) + \mu^b(\mathbf{x}) + \mu^e(\mathbf{x})] dV. \tag{1.11}$$
The differential form of (1.11) is
$$\begin{aligned}
\varepsilon_0 E_{k,k} &= \mu^l(\mathbf{x}) + \mu^b(\mathbf{x}) + \mu^e(\mathbf{x}) \\
&= -\mu^b(\mathbf{x}+\boldsymbol{\eta}) + \mu^r(\mathbf{x}) + \mu^b(\mathbf{x}) + \mu^e(\mathbf{x}) \\
&= -\mu^b(\mathbf{x}) - \mu^b_{,i}(\mathbf{x})\eta_i + \mu^r(\mathbf{x}) + \mu^b(\mathbf{x}) + \mu^e(\mathbf{x}) \\
&= -[\mu^b(\mathbf{x})\eta_i]_{,i} + \mu^b(\mathbf{x})\eta_{i,i} + \mu^r(\mathbf{x}) + \mu^e(\mathbf{x}) \\
&= -P_{i,i} + \mu^r(\mathbf{x}) + \mu^e(\mathbf{x}) = -P_{i,i} + \mu.
\end{aligned} \tag{1.12}$$
(1.12) can be written as
$$(\varepsilon_0 E_i + P_i)_{,i} = D_{i,i} = \mu, \tag{1.13}$$
where we have introduced the electric displacement vector **D** through
$$D_i = \varepsilon_0 E_i + P_i. \tag{1.14}$$

## 2. Integral Balance Laws

The relevant balance laws from physics are as follows. For the electromagnetic fields, we have Maxwell's equations
$$\oint_S \mathbf{n} \cdot \mathbf{D} dS = \int_V \mu dV, \tag{2.1}$$
$$\oint_S \mathbf{n} \cdot \mathbf{B} dS = 0, \tag{2.2}$$
$$\oint_C \mathbf{E} \cdot d\mathbf{x} + \frac{\partial}{\partial t} \int_S \mathbf{n} \cdot \mathbf{B} dS = 0, \tag{2.3}$$
$$\oint_C \mathbf{H} \cdot d\mathbf{x} = \frac{\partial}{\partial t} \int_S \mathbf{n} \cdot \mathbf{D} dS + \int_S \mathbf{n} \cdot \mathbf{J} dS, \tag{2.4}$$
and the conservation of charge
$$\frac{\partial}{\partial t} \int_V \mu dV = -\oint_S \mathbf{n} \cdot \mathbf{J} dS. \tag{2.5}$$
Various quasistatic approximations [17,18] of Maxwell's equations may be introduced when proper. The linear momentum equation for the spin continuum is
$$\int_V \left( \mathbf{f}^M + \mathbf{f}^L \right) dV = 0. \tag{2.6}$$
The angular momentum equation of the spin continuum about **x**=0 is
$$\frac{\partial}{\partial t} \int_V \frac{\mathbf{M}}{\gamma} dV = \int_V \mathbf{x} \times (\mathbf{f}^M + \mathbf{f}^L) dV + \int_S \mathbf{M} \times \mathbf{F} dS + \int_V \mathbf{M} \times \left( \mathbf{B} + \mathbf{B}^L \right) dV. \tag{2.7}$$
The energy equation and the second law of thermodynamics for the combined continuum of four are
$$\frac{\partial}{\partial t} \int_V U dV = \int_S \left( \mathbf{F} \cdot \frac{\partial \mathbf{M}}{\partial t} - \mathbf{n} \cdot \mathbf{q} \right) dS + \int_V \left( \mathbf{E} \cdot \frac{\partial \mathbf{P}}{\partial t} - \mathbf{M} \cdot \frac{\partial \mathbf{B}}{\partial t} + \mathbf{J} \cdot \mathbf{E} + R \right) dV, \tag{2.8}$$
$$\frac{\partial}{\partial t} \int_V \eta dV \geq \int_V \frac{R}{\theta} dV - \int_S \frac{\mathbf{q} \cdot \mathbf{n}}{\theta} dS. \tag{2.9}$$
The kinetic energy associated with the angular momentum of the spin continuum has been omitted because its time derivative vanishes on account of (2) of [10].

## 3. Differential Balance Laws

The differential forms of (2.1)-(2.5) are

$$\nabla \cdot \mathbf{D} = \mu, \tag{3.1}$$

$$\nabla \cdot \mathbf{B} = 0, \tag{3.2}$$

$$\nabla \times \mathbf{E} + \frac{\partial \mathbf{B}}{\partial t} = 0, \tag{3.3}$$

$$\nabla \times \mathbf{H} = \frac{\partial \mathbf{D}}{\partial t} + \mathbf{J}, \tag{3.4}$$

$$\frac{\partial \mu}{\partial t} + \nabla \cdot \mathbf{J} = 0. \tag{3.5}$$

We note that (3.5) is implied by (3.1) and (3.4). Since the spin continuum is massless, the linear momentum equation in (2.6) simply leads to $\mathbf{f}^M + \mathbf{f}^L = 0$. For the angular momentum equation of the spin continuum in (2.7), we let [1]

$$\mathbf{F} = -\mathbf{n} \cdot \mathbf{A}. \tag{3.6}$$

(1.6) requires that

$$\mathbf{A} \cdot \mathbf{M} = 0. \tag{3.7}$$

The differential form of the angular momentum equation in (2.7) is [10]

$$\frac{1}{\gamma} \frac{\partial M_i}{\partial t} = \varepsilon_{ijk} M_j \left( -A_{lk,l} + B_k + B_k^L \right), \tag{3.8}$$

where

$$A_{lk} M_{j,l} = A_{lj} M_{k,l}, \tag{3.9}$$

has been assumed. The differential forms of (2.8) and (2.9) can be obtained as:

$$\frac{\partial U}{\partial t} = -A_{ij,i} \left( \frac{\partial M_j}{\partial t} \right) - A_{ij} \left( \frac{\partial M_j}{\partial t} \right)_{,i} + E_j \frac{\partial P_j}{\partial t} - M_j \frac{\partial B_j}{\partial t} + J_i E_i + R - q_{i,i}, \tag{3.10}$$

$$\frac{\partial \eta}{\partial t} \geq \frac{R}{\theta} - \left( \frac{q_i}{\theta} \right)_{,i}. \tag{3.11}$$

## 4. Constitutive Relations

With the use of (34) of [10], the energy equation in (3.10) becomes

$$\frac{\partial U}{\partial t} = -B_j \frac{\partial M_j}{\partial t} - B_j^L \frac{\partial M_j}{\partial t} - A_{ij} \left( \frac{\partial M_j}{\partial t} \right)_{,i} + E_j \frac{\partial P_j}{\partial t} - M_j \frac{\partial B_j}{\partial t} + J_i E_i + R - q_{i,i}. \tag{4.1}$$

We introduce a free energy $F$ through the Legendre transform:

$$F = U - E_i P_i + B_i M_i - \theta \eta. \tag{4.2}$$

Then (4.1) takes the following form:

$$\frac{\partial F}{\partial t} + \frac{\partial \theta}{\partial t} \eta + \theta \frac{\partial \eta}{\partial t} = -A_{ij} \left( \frac{\partial M_j}{\partial t} \right)_{,i} - P_j \frac{\partial E_j}{\partial t} - B_j^L \frac{\partial M_j}{\partial t} + J_i E_i + R - q_{i,i}. \tag{4.3}$$

Eliminating $R$ from (3.11) and (4.3), we obtain the following inequality (Clausius–Duhem):

$$-\left( \frac{\partial F}{\partial t} + \frac{\partial \theta}{\partial t} \eta \right) - A_{ij} \left( \frac{\partial M_j}{\partial t} \right)_{,i} - P_j \frac{\partial E_j}{\partial t} - B_j^L \frac{\partial M_j}{\partial t} - \frac{q_i}{\theta} \theta_{,i} + J_i E_i \geq 0. \tag{4.4}$$

For constitutive relations we break $\mathbf{P}$ and $\mathbf{B}^L$ into reversible and dissipative parts as follows:

$$\mathbf{P} = \mathbf{P}^R\left(E_i; M_i; M_{j,i}; \theta\right) + \mathbf{P}^D\left(E_i; M_i; M_{j,i}; \theta; \theta_{,j}; \dot{E}_i; \dot{M}_i\right),$$
$$\mathbf{B}^L = {}^R\mathbf{B}^L\left(E_i; M_i; M_{j,i}; \theta\right) + {}^D\mathbf{B}^L\left(E_i; M_i; M_{j,i}; \theta; \theta_{,j}; \dot{E}_i; \dot{M}_i\right), \qquad (4.5)$$
$$\mathbf{q} = \mathbf{q}\left(E_i; M_i; M_{j,i}; \theta; \theta_{,j}; \dot{E}_i; \dot{M}_i\right), \quad \mathbf{J} = \mathbf{J}\left(E_i; M_i; M_{j,i}; \theta; \theta_{,j}; \dot{E}_i; \dot{M}_i\right).$$

The reversible parts of (4.5) are chosen to satisfy

$$\frac{\partial F}{\partial t} = -P_j^R \frac{\partial E_j}{\partial t} - {}^R B_j^L \frac{\partial M_j}{\partial t} - A_{ij}\left(\frac{\partial M_j}{\partial t}\right)_{,i} - \eta \frac{\partial \theta}{\partial t}. \qquad (4.6)$$

Then the energy equation in (4.3) and the Clausius–Duhem inequality in (4.4) become

$$\theta \frac{\partial \eta}{\partial t} = -P_j^D \frac{\partial E_j}{\partial t} - {}^D B_j^L \frac{\partial M_j}{\partial t} + J_i E_i + R - q_{i,i}, \qquad (4.7)$$

$$-P_j^D \frac{\partial E_j}{\partial t} - {}^D B_j^L \frac{\partial M_j}{\partial t} - \frac{q_i}{\theta}\theta_{,i} + J_i E_i \geq 0. \qquad (4.8)$$

(4.7) is the dissipation equation. With

$$\left(\frac{\partial M_j}{\partial t}\right)_{,i} = \frac{\partial}{\partial t}(M_{j,i}), \qquad (4.9)$$

we write (4.6) as

$$\frac{\partial F}{\partial t} = -P_j^R \frac{\partial E_j}{\partial t} - {}^R B_j^L \frac{\partial M_j}{\partial t} - A_{ij}\frac{\partial}{\partial t}(M_{j,i}) - \eta \frac{\partial \theta}{\partial t}. \qquad (4.10)$$

According to (4.10), we let

$$F = F\left(E_i; M_i; M_{j,i}; \theta\right). \qquad (4.11)$$

Then

$$\frac{\partial F}{\partial t} = \frac{\partial F}{\partial E_i}\frac{\partial E_i}{\partial t} + \frac{\partial F}{\partial M_i}\frac{\partial M_i}{\partial t} + \frac{\partial F}{\partial (M_{j,i})}\frac{\partial}{\partial t}(M_{j,i}) + \frac{\partial F}{\partial \theta}\frac{\partial \theta}{\partial t}. \qquad (4.12)$$

We substitute (4.12) into (4.10). At the same time we use Lagrange multipliers $\lambda$ and $L_i$ to include the constrains in (3)$_{1,3}$ of [10]. This yields

$$\frac{\partial F}{\partial E_i}\frac{\partial E_i}{\partial t} + \frac{\partial F}{\partial M_i}\frac{\partial M_i}{\partial t} + \frac{\partial F}{\partial (M_{j,i})}\frac{\partial}{\partial t}(M_{j,i}) + \frac{\partial F}{\partial \theta}\frac{\partial \theta}{\partial t}$$
$$= -A_{ij}\frac{\partial}{\partial t}(M_{j,i}) - P_j^R \frac{\partial E_j}{\partial t} - {}^R B_j^L \frac{\partial M_j}{\partial t} - \eta \frac{\partial \theta}{\partial t} + \lambda M_i \frac{\partial M_i}{\partial t} + L_i\left[M_{j,i}\frac{\partial M_j}{\partial t} + M_j \frac{\partial}{\partial t}(M_{j,i})\right], \qquad (4.13)$$

which can be rearranged into

$$-\left(\frac{\partial F}{\partial \theta} + \eta\right)\frac{\partial \theta}{\partial t} - \left(P_i^R + \frac{\partial F}{\partial E_i}\right)\frac{\partial E_i}{\partial t}$$
$$-\left({}^R B_i^L - \lambda M_i - L_j M_{i,j} + \frac{\partial F}{\partial M_i}\right)\frac{\partial M_i}{\partial t} - \left[A_{ij} - L_i M_j + \frac{\partial F}{\partial (M_{j,i})}\right]\frac{\partial}{\partial t}(M_{j,i}) = 0. \qquad (4.14)$$

(4.14) implies the following reversible constitutive relations:

$$ {}^R B_i^L = -\frac{\partial F}{\partial M_i} + \lambda M_i + L_m M_{i,m},$$
$$A_{ij} = -\frac{\partial F}{\partial (M_{j,i})} + L_i M_j, \qquad (4.15)$$
$$P_i^R = -\frac{\partial F}{\partial E_i}, \quad \eta = -\frac{\partial F}{\partial \theta}.$$

From (3.7) and (4.15)$_2$, we obtain

$$A_{ij}M_j = -\frac{\partial F}{\partial(M_{j,i})}M_j + L_i M_j M_j = 0. \tag{4.16}$$

(4.16) leads to

$$L_i = \frac{1}{M_s^2}\frac{\partial F}{\partial(M_{k,i})}M_k. \tag{4.17}$$

From (1.7) we have

$$\mathbf{B}^L \cdot \mathbf{M} = {}^R\mathbf{B}^L \cdot \mathbf{M} + {}^D\mathbf{B}^L \cdot \mathbf{M} = 0. \tag{4.18}$$

As a sufficient condition of (4.18), we require that

$${}^R\mathbf{B}^L \cdot \mathbf{M} = 0, \quad {}^D\mathbf{B}^L \cdot \mathbf{M} = 0. \tag{4.19}$$

From (4.15)$_1$ and (4.19)$_1$,

$${}^R B_i^L M_i = -\frac{\partial F}{\partial M_i}M_i + \lambda M_i M_i + L_j M_i M_{i,j} = -\frac{\partial F}{\partial M_i}M_i + \lambda M_i M_i = 0, \tag{4.20}$$

where (3)$_2$ of [10] has been used. (4.20) results in

$$\lambda = \frac{1}{M_s^2}\frac{\partial F}{\partial M_k}M_k. \tag{4.21}$$

Using the expressions of $\lambda$ and $L_i$ in (4.17) and (4.21), we rewrite (4.15)$_{1,2}$ as

$$\begin{aligned}{}^R B_i^L &= -\frac{\partial F}{\partial M_i} + \frac{1}{M_s^2}\frac{\partial F}{\partial M_k}M_k M_i + \frac{1}{M_s^2}\frac{\partial F}{\partial(M_{k,j})}M_k M_{i,j}, \\ A_{ij} &= -\frac{\partial F}{\partial(M_{j,i})} + \frac{1}{M_s^2}\frac{\partial F}{\partial(M_{k,i})}M_k M_j,\end{aligned} \tag{4.22}$$

where $\mathbf{A}$ is restricted by (3.9). The dissipative constitutive relations may be taken as

$${}^D B_i^L = -\beta_{ij}\frac{\partial M_j}{\partial t}, \quad \beta_{ij}\frac{\partial M_i}{\partial t}\frac{\partial M_j}{\partial t} \geq 0, \tag{4.23}$$

$$P_i^D = -\alpha_{ij}\frac{\partial E_j}{\partial t}, \quad \alpha_{ij}\frac{\partial E_i}{\partial t}\frac{\partial E_j}{\partial t} \geq 0, \tag{4.24}$$

$$q_i = -\kappa_{ij}\theta_{,j}, \quad \kappa_{ij}\theta_{,i}\theta_{,j} \geq 0, \tag{4.25}$$

$$J_i = \sigma_{ij}E_j, \quad \sigma_{ij}E_i E_j \geq 0, \tag{4.26}$$

which satisfy (4.8) and are still restricted by (4.19)$_2$.

## 5. Summary of Equations

In summary, the field equations are

$$\nabla \cdot \mathbf{D} = \mu \tag{5.1}$$

$$\nabla \cdot \mathbf{B} = 0, \tag{5.2}$$

$$\nabla \times \mathbf{E} = -\frac{\partial \mathbf{B}}{\partial t}, \tag{5.3}$$

$$\nabla \times \mathbf{H} = \mathbf{J} + \frac{\partial \mathbf{D}}{\partial t}, \tag{5.4}$$

$$\frac{\partial \mu}{\partial t} + \nabla \cdot \mathbf{J} = 0, \tag{5.5}$$

$$\frac{1}{\gamma}\frac{\partial M_i}{\partial t} = \varepsilon_{ijk}M_j\left(-A_{lk,l} + B_k + B_k^L\right), \tag{5.6}$$

$$\theta\frac{\partial \eta}{\partial t} = -P_j^D\frac{\partial E_j}{\partial t} - {}^D B_j^L\frac{\partial M_j}{\partial t} + J_i E_i + R - q_{i,i}. \tag{5.7}$$

For constitutive relations, we have

$$\mathbf{P} = \mathbf{P}^R(E_i; M_i; M_{j,i}; \theta) + \mathbf{P}^D(E_i; M_i; M_{j,i}; \theta; \theta_{,j}; \dot{E}_i; \dot{M}_i),$$
$$\mathbf{B}^L = {}^R\mathbf{B}^L(E_i; M_i; M_{j,i}; \theta) + {}^D\mathbf{B}^L(E_i; M_i; M_{j,i}; \theta; \theta_{,j}; \dot{E}_i; \dot{M}_i), \tag{5.8}$$
$$\mathbf{q} = \mathbf{q}(E_i; M_i; M_{j,i}; \theta; \theta_{,j}; \dot{E}_i; \dot{M}_i), \quad \mathbf{J} = \mathbf{J}(E_i; M_i; M_{j,i}; \theta; \theta_{,j}; \dot{E}_i; \dot{M}_i).$$

$$F = F(E_j; M_i; M_{j,i}; \theta), \tag{5.9}$$

$$
{}^R B_i^L = -\frac{\partial F}{\partial M_i} + \frac{1}{M_s^2}\frac{\partial F}{\partial M_k}M_k M_i + \frac{1}{M_s^2}\frac{\partial F}{\partial (M_{k,j})}M_k M_{i,j},
$$

$$
A_{ij} = -\frac{\partial F}{\partial (M_{j,i})} + \frac{1}{M_s^2}\frac{\partial F}{\partial (M_{k,i})}M_k M_j, \tag{5.10}
$$

$$
P_i^R = -\frac{\partial F}{\partial E_i}, \quad \eta = -\frac{\partial F}{\partial \theta},
$$

which are restricted by

$$A_{lk} M_{j,l} = A_{lj} M_{k,l}, \tag{5.11}$$

$$-P_j^D \frac{\partial E_j}{\partial t} - {}^D B_j^L \frac{\partial M_j}{\partial t} - \frac{q_i}{\theta}\theta_{,i} + J_i E_i \geq 0, \tag{5.12}$$

$${}^D \mathbf{B}^L \cdot \mathbf{M} = 0. \tag{5.13}$$

In addition, we have

$$\mathbf{D} = \varepsilon_0 \mathbf{E} + \mathbf{P}, \quad \mathbf{H} = \frac{\mathbf{B}}{\mu_0} - \mathbf{M}. \tag{5.14}$$

On a boundary surface with an outward unit normal **n**, in addition to the usual boundary conditions for the electromagnetic fields in a conductor, possible boundary conditions for the thermal and **M** fields are [1]

$$\theta \quad \text{or} \quad \mathbf{n} \cdot \mathbf{q},$$
$$\delta\boldsymbol{\theta} \quad \text{or} \quad \mathbf{n} \cdot \mathbf{A} \times \mathbf{M}, \tag{5.15}$$

where $\theta$ is the absolute temperature and $\delta\boldsymbol{\theta}$ is the angular displacement of **M** [1,10].

## 6. Conclusions

The four-continuum model can describe the basic behaviors of polarization, magnetization and electrical conduction in rigid ferromagnetic conductors. The model can be used to construct a macroscopic theory of ferromagnetic conductors in a systematic way. The theory constructed includes the LLG equation for damped spin waves as one of a system of equations. It also includes thermal and various dissipative effects.